\documentclass[aps,prl,superscriptaddress,notitlepage,twocolumn,showpacs]{revtex4-1}
\usepackage{amssymb, amsmath, amsfonts}
\usepackage{graphicx}
\usepackage[hidelinks]{hyperref}

\renewcommand{\Re}{\operatorname{Re}}
\renewcommand{\Im}{\operatorname{Im}}

\newcommand{\dca}{\Delta_\text{ca}}
\newcommand{\dpc}{\Delta_\text{pc}}
\newcommand{\starkScalar}{\alpha_0}
\newcommand{\starkVector}{\alpha_1}
\newcommand{\larmorW}{\omega_L}
\newcommand{\cavityW}{\omega_c}
\newcommand{\atomN}{N}
\newcommand{\nbar}{\bar{n}}
\newcommand{\Fx}{\hat{F}_x}
\newcommand{\Fz}{\hat{F}_z}
\newcommand{\Bopt}{\mathbf{B}_\text{opt}}
\newcommand{\Gopt}{\Gamma_\text{opt}}
\newcommand{\dWopt}{\delta \omega_\text{opt}}

\begin{document}

\title{Cavity-assisted measurement and coherent control of collective atomic spin oscillators}

\author{Jonathan Kohler}
\email{jkohler@berkeley.edu}
\affiliation{Department of Physics, University of California, Berkeley, CA 94720, USA}
\author{Nicolas Spethmann}
\affiliation{Department of Physics, University of California, Berkeley, CA 94720, USA}
\affiliation{Department of Physics and Research Center OPTIMAS, Technische Universit\"at Kaiserslautern, 67663 Kaiserslautern, Germany}
\author{Sydney Schreppler}
\affiliation{Department of Physics, University of California, Berkeley, CA 94720, USA}
\author{Dan M.\ Stamper-Kurn}
\email{dmsk@berkeley.edu}
\affiliation{Department of Physics, University of California, Berkeley, CA 94720, USA}
\affiliation{Materials Sciences Division, Lawrence Berkeley National Laboratory, Berkeley, CA  94720, USA}
\date{\today}

\begin{abstract}
We demonstrate continuous measurement and coherent control of the collective spin of an atomic ensemble undergoing Larmor precession in a high-finesse optical cavity.  The coupling of the precessing spin to the cavity field yields phenomena similar to those observed in cavity optomechanics, including cavity amplification, damping, and optical spring shifts.  These effects arise from autonomous optical feedback onto the atomic spin dynamics, conditioned by the cavity spectrum.  We use this feedback to stabilize the spin in either its high- or low-energy state, where, in equilibrium with measurement back-action heating, it achieves a steady-state temperature, indicated by an asymmetry between the Stokes and anti-Stokes scattering rates.  For sufficiently large Larmor frequency, such feedback stabilizes the spin ensemble in a nearly pure quantum state, in spite of continuous measurement by the cavity field.
\end{abstract}


\maketitle

Quantum systems are invariably perturbed by measurement. For a repeated or continuous measurement, the measurement back-action perturbing the system generally adds noise to the subsequent measurement record, reducing the precision of state estimation and measurement sensitivity of external forces \cite{Caves1980}.
However, this additional noise carries useful information about the measurement-induced perturbation, allowing for feedback that suppresses the effects of back-action and controls the evolution of the quantum system \cite{Wiseman1994}.  Such feedback may be either measurement-based, where the quantum system is driven externally based on classical information gleaned from quantum measurements \cite{Kubanek2009,Sayrin2011,Vijay2012,Bushev2013,Behbood2013,Wilson2015}, or autonomous, where feedback is implemented through coherent \cite{Mabuchi2008} or dissipative \cite{Kraus2008,Krauter2011,Shankar2013} dynamics inherent to the system itself.

Atomic spin ensembles in cavity quantum electrodynamic systems have been demonstrated as prototypical examples of quantum measurement and control.
Many recent experiments with spin-cavity systems have focused on avoiding back-action to overcome standard quantum limits \cite{Braginsky1980} and to prepare squeezed states for improved metrology through quantum non-demolition \cite{Leroux2010,Sewell2012,Bohnet2014,Hosten2016,Cox2016} or back-action evading \cite{Vasilakis2015} measurements.
Here, we focus instead on continuous weak detection of non-stationary observables in order to study quantum-limited back-action and autonomous stabilization effects using the cavity's finite lifetime.

In this Letter, we report the observation of back-action effects from weak continuous measurement of the Larmor precession of an atomic spin ensemble within a driven high-finesse optical cavity.  By recirculating the light probing the spin ensemble, the cavity allows the optical modulation induced by the precessing spin to act back on the subsequent spin dynamics.  The cavity thereby conditions a feedback loop that allows the accumulated noise from measurement back-action to be suppressed and the collective spin to be stabilized near either its lowest- or highest-energy state, chosen by the detuning of the cavity probe light from the cavity resonance.  We find that, for sufficiently large Larmor frequency, such a feedback-stabilized spin oscillator remains in a nearly pure quantum state, in spite of continuous interaction with the probe field.

Consider an ensemble of $\atomN$ identical atoms in their electronic ground state, each with spin $f$.  The dimensionless collective spin, $\hat{\mathbf{F}}$, is sensed optically via the circular birefringence of the atoms.  Specifically, a beam of circularly polarized light propagating along the $\mathbf{z}$ axis acquires a phase shift proportional to $\Fz$, the projection of the collective spin onto the axis of optical angular momentum.

\begin{figure}[t!]
	\includegraphics[width=3.4in]{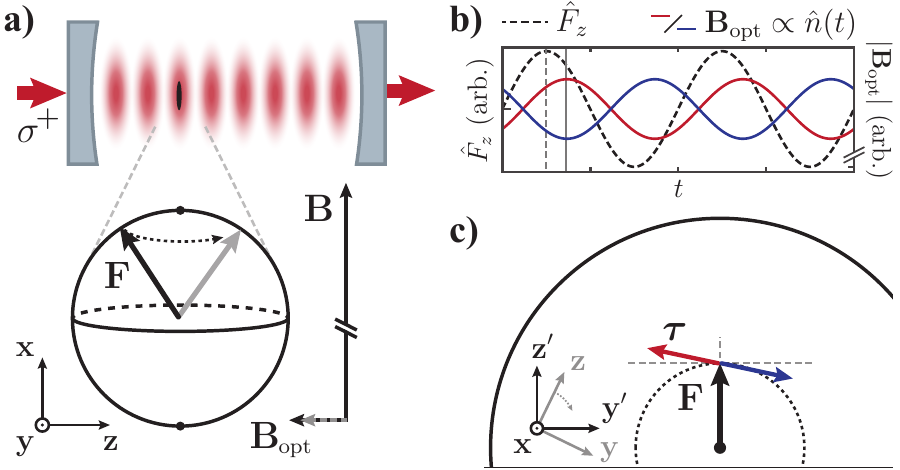}
	\caption{
		\label{fig:schematic} (color online) \textbf{(a)} Schematic of the experimental system.
		Atoms trapped at an anti-node of the cavity field experience an effective magnetic field $\Bopt$ due to interaction with circularly polarized light.  A large external magnetic field $\mathbf{B}$ along $\mathbf{x}$ defines the Larmor frequency $\larmorW$ and high- and low-energy stable poles of the dynamics.  \textbf{(b)} The oscillating transverse spin component $\Fz$ couples to the cavity field, causing amplitude modulation for a cavity driven either above (blue) or below (red) resonance.  The optical response is delayed due to the finite cavity linewidth  $\kappa$.  \textbf{(c)} The average torque $\mathbf{\tau}$ acting on the spin due to this optical feedback, in a rotating frame. The in-phase modulation generates torque tangential to the spin trajectory, shifting the Larmor precession frequency. The out-of-phase component generates torque perpendicular to this trajectory, causing spin nutation toward or away from the poles.	
}
\end{figure}

An applied magnetic field $\mathbf{B}$ induces the Hamiltonian $\mathcal{H}_B = -\hbar \gamma \mathbf{B} \cdot \hat{\mathbf{F}}$, where $\gamma$ is the  atomic gyromagnetic ratio. When $\mathbf{B}$ is oriented away from $\mathbf{z}$ -- say, along $\mathbf{x}$ (see Fig.\ \ref{fig:schematic}a) -- the light measures oscillations of one transverse spin component, allowing real-time observation of Larmor precession of the collective spin.  The light acts back on the precessing spin as an effective magnetic field $\Bopt \parallel \mathbf{z}$, proportional to the probe light intensity.  Measurement back-action arises from quantum fluctuations of the light intensity, and, hence, of $\Bopt$.  The resulting fluctuations in the torque produced by $\Bopt$ affect both the phase of Larmor precession and the energy of the spin ensemble.

Inside a high-finesse cavity, optical fluctuations persist for a finite life-time, allowing them to coherently alter the subsequent spin dynamics.  In particular, when the cavity is driven off-resonance, the precessing spin induces amplitude modulation of the probe field.  This amplitude modulation acts back on the precessing spin through the effective magnetic field $\Bopt$, which varies synchronously with the Larmor precession.  The finite cavity lifetime leads to a delay between the oscillating spin displacement $\Fz(t)$ and the modulation of the intra-cavity photon number $\hat{n}(t)$ (Fig.\ \ref{fig:schematic}b).

The modulation of $\Bopt$ has two effects, which become clear when considered in a frame co-rotating with the precessing spin around $\mathbf{B}$ \cite{Brahms2010}.  Under the rotating-wave approximation, the modulation component in-phase with the oscillation of $\Fz$ generates a net torque tangential to the instantaneous trajectory of the precessing spin  (Fig.\ \ref{fig:schematic}c), shifting the effective Larmor frequency.  The out-of-phase component generates an average torque acting perpendicular to this trajectory, causing the atomic spin to nutate toward either the low-energy or the high-energy pole, depending on the sign of the optical modulation.  The relative phase between the spin precession and the cavity field modulation is controlled by the probe detuning from cavity resonance $\dpc$, and by the ratio of $\larmorW$ to the cavity half-linewidth $\kappa = 2\pi \times 1.82$ MHz.

This Letter extends concepts from cavity optomechanics \cite{Aspelmeyer2014} to the cavity-based detection and control of collective spin modes, which was discussed theoretically in Ref.\ \cite{Brahms2010}, and also recently in terms of solid-state cavity opto-magnonics \cite{Osada2016, Liu2016, Kusminskiy2016}.  The autonomous feedback driving the collective spin toward or away from either the low- or high-energy states is analogous to cavity-induced damping\ \cite{Arcizet2006,Gigan2006,Schliesser2006,Thompson2008} or amplification \cite{Kippenberg2005} of motion.  We also observe feedback-induced shifts in the Larmor frequency that correspond to the optical spring effect \cite{Sheard2004,Corbitt2006}.  Furthermore, similar to experiments in cavity optomechanics \cite{Brahms2012,Safavi-Naeini2012}, we observe asymmetry between sidebands generated by optical Stokes and anti-Stokes processes, allowing us to characterize the effective spin temperature reached by the balance of coherent and incoherent back-action effects.

Our experiments are performed on gases of about $\atomN = 3500$ $^{87}$Rb atoms placed within a Fabry-P\'erot cavity \cite{Purdy2010}.  The atoms are evaporatively cooled to around 5 $\mu$K and trapped within a couple adjacent sites of a 64 $\mu$K deep, one-dimensional, spin-independent optical lattice, created by driving a $\text{TEM}_{00}$ mode of the cavity with light at a wavelength of 860 nm.
A uniform magnetic field $\mathbf{B}$ is applied along $\mathbf{x}$,  transverse to the cavity axis (Fig.\ \ref{fig:schematic}a), and its magnitude sets the Larmor frequency $\larmorW = |\gamma \mathbf{B}|$, which we vary from $\larmorW/2\pi = 100$ kHz to several MHz.
The atoms are prepared initially in the $\left| f=2, m_F = +2\right>$ hyperfine state (with the quantization axis along $\mathbf{B}$).  The collective atomic spin, with total spin $F = 2 N \sim 7000$, is thus prepared in its highest energy state,  since $\gamma<0$ for the $f=2$ hyperfine level of $^{87}$Rb.
Alternately, the ensemble can be rotated to its lowest-energy spin state by applying a $\pi$ pulse to the atoms using a radio-frequency magnetic field.

We probe the ensemble through its influence on a different TEM$_{00}$ mode of the cavity, whose resonance frequency $\cavityW$ is detuned by $\dca/2\pi = -42$ GHz from the $^{87}$Rb D2 transition (with wavelength $780$ nm).  At this large detuning, the atom-cavity interaction is predominately dispersive.  Because the detuning is also much larger than the excited-state hyperfine splitting, the interaction is dominated by scalar and vector terms, and higher-order tensor interactions are negligible.

The cavity supports modes of two independent polarizations at frequencies near $\cavityW$. Driving the cavity with a weak coherent beam of one circular polarization of light, with helicity along either $\pm \mathbf{z}$, and neglecting the small linear birefringence of our cavity\ \footnote{The cavity's linear birefringence splits the resonance frequencies of two orthogonal linear polarizations by $\Delta_b / 2 \pi = 1.2$ MHz.  This leads to a small mixing between circular polarizations, with the lowest-order correction in photon number of the un-driven mode proportional to $(\Delta_b / 2 \kappa)^2 = 0.11$, which was shown to have a negligible effect in numerical simulations.}, the resulting evolution is described by the effective system Hamiltonian
\begin{equation}
\label{eq:hamiltonian}
\mathcal{H} =
\hbar \cavityW \hat{a}^\dagger \hat{a} + \hbar \larmorW \Fx +  \frac{\hbar g_0^2}{\dca} \hat{a}^\dagger \hat{a} (\starkScalar \atomN \pm \starkVector \Fz)\text{,}
\end{equation}
obtained by adiabatically eliminating the atomic excited states.  Here, $\hat{a}$ is the creation operator for cavity photons, and the coefficients $\starkScalar=2/3$ and $\starkVector=1/6$ describe the relative strength of the scalar and vector parts of the ac Stark shift, determined by summing over all excited-state hyperfine levels.
The position-averaged vacuum Rabi coupling $g_0/2\pi = 13$ MHz is evaluated by considering the geometry of the cavity mode \cite{Purdy2010}.  In this work, atoms are trapped near antinodes of the cavity probe field to minimize linear optomechanical effects. The coupling of the cavity field to the environment is further described using the standard input-output formalism, and the modulated probe field transmitted through the cavity is measured with an optical heterodyne detector, with overall cavity photon detection efficiency $\epsilon=0.12$.

\begin{figure}[!t]
	\centering
	\includegraphics{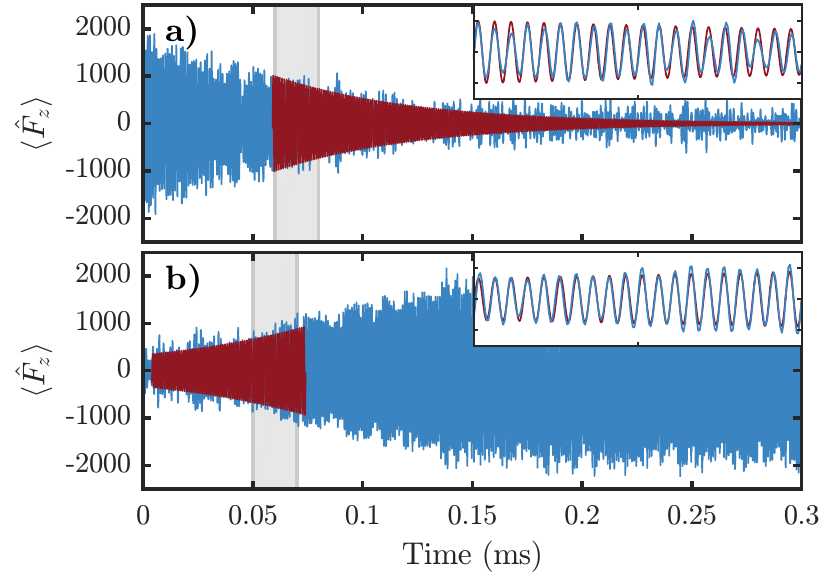}
	\caption{
		\label{fig:timedomain} (color online) Coherent damping and amplification of Larmor precession of a spin ensemble with $\larmorW/2\pi=1.0$ MHz, observed in the phase modulation of transmitted light, averaged over 30-40 repetitions (blue).  Cavity probe light ($\dpc/2\pi = 1.0$ MHz, $\nbar=4$ average intracavity photons) drives the spin toward the high-energy pole.
		\textbf{(a)} Larmor precession of a spin ensemble, displaced from the high-energy pole by a $\pi/10$ rf pulse, coherently damps back to the pole at a rate $\Gopt/2\pi=4.9 \pm 0.2$ kHz.
		\textbf{(b)} A spin prepared near the low-energy pole, by application of a near $\pi$-pulse, is coherently amplified away at a rate $\Gopt/2\pi=-4.6 \pm 0.4$ kHz.
		Exponential rates are extracted by simultaneous fits (red) of both amplitude and phase quadratures. Insets show the harmonic nature of the Larmor precession signal and quality of fit in the highlighted regions.	
		The finite cavity linewidth causes the observed signal to saturate at around $2000$.}
\end{figure}

The spin dynamics imprinted on the cavity output field are observed in the demodulated heterodyne signal. For example, in Fig.\ \ref{fig:timedomain}, we compare the evolution of spins prepared near either the high- or low-energy poles when the cavity is driven by a blue-detuned ($\dpc>0$) probe.  In both cases, the probe drives the spin toward the high-energy pole.  For a spin prepared initially near the high-energy pole, cavity back-action coherently damps the Larmor precession amplitude, analogous to cavity optomechanical cooling.  In comparison, the Larmor precession of a spin prepared near the low-energy pole is coherently amplified, analogous to regenerative optomechanical amplification.  At longer times (not shown in the Figure), in the latter case, the ensemble's spin nutates past the equator of the Bloch sphere and also damps back to the high-energy pole.  If instead we drive the cavity with red-detuned probe light ($\Delta_{pc}<0$), we observe similar behavior, with the collective spin instead driven toward and stabilized at the low-energy pole.

\begin{figure}[!b]
	\includegraphics{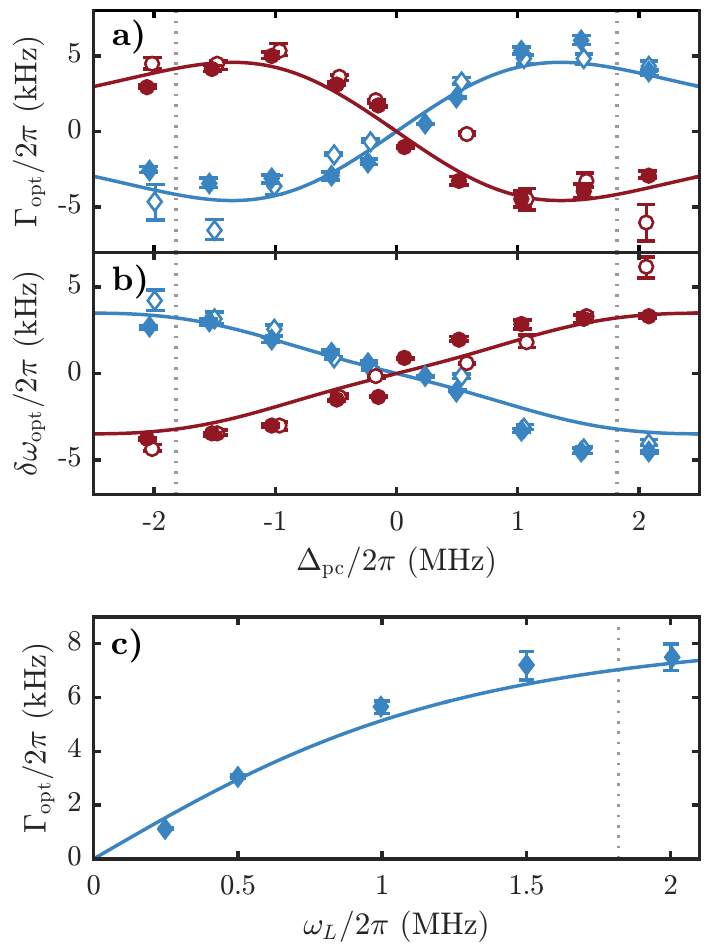}
	\caption{\label{fig:damping} (color online) \textbf{(a)} Optical damping rates and \textbf{(b)} frequency shifts of Larmor precession as a function of probe detuning $\dpc$, with fixed intracavity intensity $\nbar=4$ and $\larmorW/2\pi=1.0$ MHz.  Diamonds (blue) label results for an ensemble initially prepared near the high-energy pole, and circles (red) for an ensemble initially near the low-energy pole.
		Measurements repeated with either $\sigma^{+}$ (solid symbols) or $\sigma^{-}$ (open symbols) circularly polarized light demonstrate independence of optical helicity.
		\textbf{(c)} Peak damping rate as a function of Larmor frequency.  Dotted vertical lines mark the position of the cavity half-linewidth $\kappa$.  All theory lines are plotted with no free parameters.
		Error bars reflect statistical uncertainties from the fits.  Additional systematic errors in the probe frequency stability and initial spin state preparation predominately affect data at small probe detuning.  }
\end{figure}

The light-induced driving of a spin ensemble to either the low- ($\dpc<0$) or high-energy ($\dpc>0$) pole is reminiscent of optical pumping \cite{Happer1972}.
However, unlike optical pumping, the dynamics in our experiment cannot spontaneously generate spin polarization.
In addition, while optical pumping uses circularly polarized light to pump \emph{angular momentum} into an atomic gas, the asymmetric fluctuation spectrum of the cavity optical field is used to pump \emph{energy} into the atomic system. Indeed, we confirm that the dynamics are quantitatively the same for both circular probe polarizations (Fig.\ \ref{fig:damping}a-b).

These dynamics may also be described in terms of cavity superradiance \cite{Dicke1954,Gross1979,Black2003,Slama2007,Norcia2016}.  Consider an atomic spin ensemble initialized in the low-energy spin state.  The optically driven atoms lie in a virtually excited state from which they may decay by Raman scattering into the cavity mode.  When the cavity is driven at a positive detuning, the cavity Purcell effect induces Raman emission preferentially on the Stokes sideband, creating transverse coherence in the ensemble.  Such coherence stimulates Raman scattering at an enhanced rate, driving the spins exponentially away from the low-energy pole.
 
A quantitative treatment for the coherent dynamics near both poles can be derived classically, as in the theory of linear cavity optomechanics \cite{Aspelmeyer2014}, by neglecting the quantum noise of the cavity field.  We find that the inverse susceptibility of the spin oscillator to torque $\tau(\omega)$ at frequency $\omega$ is given by $\chi_s^{-1}(\omega) = \larmorW^2 - \omega^2 + \Sigma(\omega)$, with
\begin{equation}
\label{eq:optical-susceptibility}
\Sigma (\omega) = 2 \beta \nbar g_s^2
\Bigl(
	\frac{\larmorW}{\omega - \dpc + i \kappa} - \frac{\larmorW}{\omega + \dpc + i \kappa}
\Bigr)\text{,}
\end{equation}
where $\beta= \operatorname{sgn}(\Fx)$ labels the high-energy ($\beta=+1$) or low-energy ($\beta=-1$) state, $\nbar$ is the average photon number in the cavity mode, and $g_s = \starkVector g_0^2 \sqrt{F/2} / \dca $ describes the relevant single-photon, single-excitation coupling \cite{Brahms2010}.
The real and imaginary parts of Eq.\ \ref{eq:optical-susceptibility} describe the cavity-induced  Larmor frequency shift and the exponential damping or amplification rate, given by $\dWopt = \Re[\Sigma(\larmorW)]/ 2\larmorW$ and $\Gopt = -\Im \big[\Sigma(\larmorW) \big]/\larmorW$, respectively.

Averaging the coherent heterodyne signal from up to 40 repetitions of the experiment to reduce statistical noise, we extract both $\Gopt$ and $\dWopt$ from early times in the measured transient response (Fig.\ \ref{fig:damping}a-b).  For this, we simultaneously fit both quadratures of the heterodyne signal, restricting the fit to spin dynamics near the poles (within $0.8$ rad, see Supplemental Material\ \footnote{See Supplemental Material at [SMURL], which includes Ref. \cite{Holstein1940}, for a description of the fit procedure and derivations of plotted theoretical predictions.}).

The close agreement between our measurements and the theoretical predictions (Fig.\ \ref{fig:damping}) demonstrates the close analogy between cavity optomechanics and spin optodynamics.  In addition, for a spin oscillator, $\larmorW$ can be tuned readily over a broad range.  Therefore, we can observe optodynamical effects in the transition from the unresolved- to the resolved-sideband regime.  For example, we demonstrate that the peak damping rate, measured for a range of Larmor frequencies, is maximized for $\larmorW/\kappa>1$ (Fig.\ \ref{fig:damping}c).

In addition to coherent cavity back-action, the collective spin is also subject to quantum fluctuations of $\Bopt$, which cause diffusion of the collective spin away from its stable state.  The balance between coherent and incoherent back-action effects leads the spin ensemble to achieve a steady-state temperature.  This equilibrium between cavity-assisted damping and measurement back-action is similar to that achieved in cavity cooling of mechanical oscillators \cite{Marquardt2007,Wilson-Rae2007}.  Because the atomic spins are extremely isolated from their environment, high spin optodynamical cooperativity, where the optical coupling exceeds the intrinsic damping, is achieved already at minimal optical power.  Therefore, the equilibrium spin temperature is determined solely by the quantum fluctuations of the cavity field, which are shaped by the cavity linewidth.

Due to the quantization of the collective spin oscillator, the equilibrium temperature can be determined from the asymmetry in its Raman scattering spectrum, recorded in the power spectral density (PSD) of the heterodyne signal.  The amplitude of the Stokes sideband corresponds to the energy absorption rate $\Gamma_{+}$ and the anti-Stokes sideband corresponds to the emission rate $\Gamma_{-}$.  Assuming the collective spin state to be in thermal equilibrium at temperature $T$, the ratio of these rates satisfies the detailed balance relation $\Gamma_{+}/\Gamma_{-} = \exp (\hbar \larmorW / k_B T)$.

\begin{figure}[!t]
	\centering
	\includegraphics{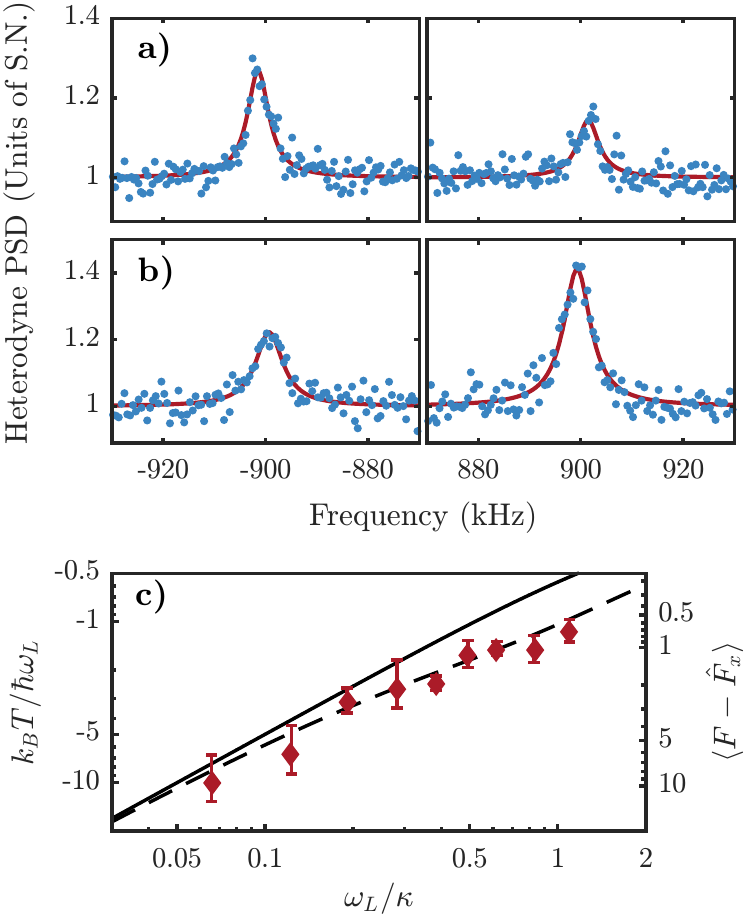}
	\caption{\label{fig:asymmetry}
		(color online) Sideband-asymmetry thermometry of a spin oscillator.
		\textbf{(a)} Stokes and anti-Stokes sidebands observed in the averaged heterodyne PSD, normalized by the shot-noise level, showing the sideband asymmetry of a low-energy spin, with $\larmorW/2\pi = 900$ kHz, in equilibrium with an optical damping tone detuned by $\dpc/2\pi = -1.5$ MHz with average photon number $\nbar_\text{damp}=5.0$, and measurement back-action from an on-resonance probe with average photon number $\nbar_\text{probe}=1.0$.
		\textbf{(b)} Inverted sideband asymmetry for a negative-temperature spin ensemble near its high-energy state, with $\dpc/2\pi = 1.5$ MHz, $\nbar_\text{damp}=4.3$, and $\nbar_\text{probe}=2.0$.
		\textbf{(c)} Equilibrium temperature (left scale) and thermal occupation (right scale) of a high-energy spin oscillator, measured by sideband asymmetry, as a function of the Larmor frequency $\larmorW$.  Lines indicate expected temperature at equilibrium between coherent damping and incoherent back-action.  The solid line indicates the ideal limit in the absence of the on-resonance probe, and the dashed line includes the additional back-action for the average probe cooperativity.  }
\end{figure}

We characterize the steady state of the atomic spin ensemble by driving the optical cavity with two coherent, primarily shot-noise limited tones: one strong tone at a detuning $\dpc=\beta \sqrt{\kappa^2 + \larmorW^2}$ that provides damping toward either the high- or low-energy pole, and a second weak tone on cavity resonance.  The asymmetry of sidebands generated on the resonant tone is observed in our heterodyne receiver (Fig.\ \ref{fig:asymmetry}a) and used to determine $T$.  Whereas a spin oscillator stabilized in its low-energy state yields the conventional Stokes/anti-Stokes sideband asymmetry, consistent with positive temperature, for a spin oscillator in the high-energy state, the sideband asymmetry is reversed, indicating a negative temperature for the spin ensemble (Fig.\ \ref{fig:asymmetry}b).  Accounting for shot-noise-driven heating of the spin oscillator from both probe tones, the observed spin temperatures agree well with the theoretical model (Fig.\ \ref{fig:asymmetry}c).  As predicted \cite{Marquardt2007,Wilson-Rae2007}, autonomous feedback cooling yields increased purity of the quantum state ($k_B |T| / \hbar \larmorW < 1$) as the system enters the resolved-sideband regime.

In summary, we have demonstrated cavity based measurement of Larmor precession of the collective spin of an atomic ensemble and coherent control via autonomous feedback. These capabilities could be used to perform quantum-limited measurements of the collective spin or to realize a phase-preserving, quantum-limited amplifier for spin states.  Furthermore, we demonstrate the ability to stabilize the ensemble in a nearly-pure quantum state with negative effective temperature.  In this condition, the system can be described as a near ground-state, negative effective-mass oscillator. If measured jointly with a mechanical oscillator, such a spin oscillator could allow continuous QND position or force measurement via coherent quantum noise cancellation \cite{Tsang2012}.

\begin{acknowledgments}
We thank L. Buchmann for helpful discussions and J. Gerber for assistance in the lab. This work was supported by the Air Force Office of Scientific Research.  N.S.\ was supported by a Marie Curie International Outgoing Fellowship, J.K.\ and S.S.\ by the U.S. Department of Defense through the National Defense Science and Engineering Graduate Fellowship program.
\end{acknowledgments}

%


\onecolumngrid

\newcommand{\PRLsep}{\vspace{1em}\noindent\makebox[\linewidth]{\resizebox{0.6\linewidth}{1pt}{$\blacklozenge$}}\bigskip}

\PRLsep

\twocolumngrid

\appendix

\renewcommand{\theequation}{S\arabic{equation}}
\renewcommand{\thefigure}{S\arabic{figure}}
\setcounter{figure}{0}
\setcounter{equation}{0}

\section*{Supplemental Material}

\section*{S.1 Damping and amplification rates}

We reproduce here some results from Ref. \cite{Brahms2010}, providing a more detailed derivation of the connection to linear optomechanical theory and a foundation for the analysis in the next section used to estimate the spin dynamics from the measured optical modulation.

To obtain an analytic solution for small-amplitude dynamics, we linearize the system Hamiltonian to describe modulations of the spin around the magnetic poles and of the cavity field about its mean value.  Transforming into a cavity frame rotating at the probe laser frequency $\omega_p$, then expanding to first order about the average cavity field $\hat{a} \rightarrow \bar{a} + \hat{a}(t)$ and spin projections $\hat{F}_z \rightarrow \bar{F}_z + \hat{F}_z(t)$ and $\hat{F}_x \rightarrow \bar{F}_x + \hat{F}_x(t)$, dropping constant terms, the linearized Hamiltonian describing interactions with circularly polarized light with helicity along $+\mathbf{z}$ is given by
\begin{multline}
\label{eq:linear-hamiltonian}
\mathcal{H} = 
- \hbar \Delta_\text{pc} \hat{a}^\dagger \hat{a}
+ \hbar \omega_L \hat{F}_x 
+ \hbar g_c \bar{a} (\hat{a}^\dagger + \hat{a}) \hat{F}_z \\
+ \hbar g_c \bar{a}^2 \hat{F}_z 
+ \hbar g_c \bar{F}_z \hat{a}^\dagger \hat{a}\text{,}
\end{multline}
where $g_c = \alpha_1 g_0^2/ \Delta_\text{ca} = 2 \pi \times 671$ Hz and $\Delta_\text{pc} = \omega_p - \omega_c^\prime$ is the probe detuning from cavity resonance, shifted by the scalar ac Stark shift $\omega_c^\prime = \omega_c + \alpha_0 N g_0^2/ \Delta_\text{ca}$.  We define the average field amplitude $\bar{a} = \sqrt{\bar{n}}$ to be real without loss of generality. This linearized approximation is valid for spin amplitudes that modulate the cavity resonance by much less than a cavity line-width, quantified by $g_c \bar{F}_z \ll \kappa$.

The average cavity intensity interacts with the atoms like an effective static magnetic field, represented in the fourth term of the above Hamiltonian.  This field rotates the static orientation of the spin slightly away from the $\mathbf{x}$ axis, creating a small average projection $\bar{F}_z \simeq ( g_c \bar{n} / \omega_L ) \bar{F}_x$, which is negligible in the limit $ g_c\bar{n} / \omega_L \ll 1$.  This rotation also causes an additional static shift to the cavity resonance, reflected in the fifth term above, which can be absorbed into the net probe detuning $\Delta_\text{pc}$.  

Neglecting these small static corrections, the equations of motion for the collective spin, obtained from the linearized Hamiltonian, are given by
\begin{subequations}
\begin{align}
\label{eq:spin-eoms}
\dot{\hat{F}}_x & = - g_c \sqrt{\bar{n}} (\hat{a}^\dagger + \hat{a}) \hat{F}_y \text{,} 
 \\
\dot{\hat{F}}_y & = -\omega_L \hat{F}_z  +  g_c \sqrt{\bar{n}} (\hat{a}^\dagger + \hat{a}) \hat{F}_x \text{,} 
 \\
\dot{\hat{F}}_z & = \omega_L \hat{F}_y
\text{.}
\end{align}
\end{subequations}
For small excitations away from the stable poles, we approximate $\hat{F}_x \approx \beta F$, where $\beta=\pm 1$ labels the high- and low-energy states, and these simplify to
\begin{equation}
\label{eq:linear-spin-eom}
\ddot{\hat{F}}_z = -\omega_L^2 \hat{F}_z + \beta g_c \sqrt{\bar{n}} \omega_L F ( \hat{a}^\dagger + \hat{a})\text{.}
\end{equation}
The equation of motion for the cavity field, including the vacuum noise at the cavity input $\hat{\xi}$ and dissipation of the cavity field, is 
\begin{equation}
\label{eq:linear-cavity-eom}
\dot{\hat{a}} = (i \Delta_\text{pc} - \kappa) \hat{a} - i g_c \sqrt{\bar{n}} \hat{F}_z + \sqrt{2 \kappa} \hat{\xi}\text{.}
\end{equation}

These equations closely resemble those describing linear optomechanics, with the spin projection along the cavity axis $\hat{F}_z$ representing the generalized position of the oscillator.  Near either pole, excitations of the collective spin away from the high- or low-energy state can be approximated as bosonic excitations of a harmonic oscillator mode $\hat{b}$\ \cite{Holstein1940}.  We define an effective unit-less spin displacement $\hat{z}_s = \hat{b}^\dagger + \hat{b} = \hat{F}_z / \Delta_\text{SQL}$, where $\Delta_\text{SQL} = \sqrt{\frac{F}{2}}$, the standard quantum limit for spin fluctuations, is the spin equivalent of the quantum zero-point motion.  Under this approximation, the analogy to optomechanics is exact, and the equations can be written as
\begin{subequations}
\label{eq:harmonic-eoms}
\begin{align}
\ddot{\hat{z}}_s & = -\omega_L^2 \hat{z}_s +  2 \beta \sqrt{\bar{n}} g_s \omega_L (\hat{a}^\dagger + \hat{a}) \text{,} \\ 
\dot{\hat{a}} &= (i \Delta - \kappa) \hat{a} - i \sqrt{\bar{n}} g_s \hat{z}_s + \sqrt{2 \kappa} \hat{\xi}\text{,}
\end{align}
\end{subequations}
by defining the analogous single-photon, single-excitation coupling $g_s = g_c \Delta_\text{SQL}$.

Semi-classical solutions for the coherent dynamics of the system can be found by dropping the vacuum input noise and considering the spin displacement in response to a unit-less torque $\tau(\omega)$ at frequency $\omega$.  The complex susceptibility of the oscillator response $\chi_s(\omega)=z_s(\omega) / \tau(\omega)$ is found by solving the Fourier transform of Equations\ \ref{eq:harmonic-eoms}, resulting in
\begin{equation}
\chi_s^{-1}(\omega) = \omega_L^2 - \omega^2 + \Sigma(\omega)\text{,}
\end{equation}
where $\Sigma(\omega)$ is given by Equation 2 in the main text.
	
\section*{S.2 Simultaneous Quadrature Fit}

To obtain an estimate of the coherent spin dynamics $\langle \hat{F}_z(t) \rangle$ from the recorded optical modulation, we solve the equation of motion for the cavity field, neglecting quantum fluctuations.  Assuming the spin oscillator's response takes the form $F_z(t) = A(t) \sin ( \omega_L t)$, where $A(t)$ is a slowly varying amplitude, the solution to Equation\ \ref{eq:linear-cavity-eom} is approximately
\begin{multline}
a(t) = \frac{i \sqrt{\bar{n}} g_c}{(\kappa - i \Delta_\text{pc})^2 + \omega_L^2} A(t) \\
 \times \Bigl[
(\kappa - i \Delta_\text{pc}) \cos \omega_L t + \omega_L \sin \omega_L t
\Bigr]\text{.}
\end{multline}

The cavity output field is determined by the boundary condition $\hat{a}_\text{in} + \hat{a}_\text{out} = \sqrt{2 \kappa} \hat{a}$, where the input field on the open port of our double-sided cavity is only vacuum noise, which we also neglect in this semi-classical treatment.  We perform a balanced heterodyne detection of this field, combining it with a local oscillator, with power $P_\text{LO}= 1.0$ mW, derived from the same source and shifted $\omega_0$ relative to the probe.  Demodulating the resulting photocurrent, the detected optical power expected in the amplitude and phase quadratures is given by
\begin{subequations}
\label{scaled-photocurrents}
\begin{align}
\label{eq:phase-quadrature}
\langle I_{Q}(t) \rangle & = B \sqrt{\kappa^2 + \omega_L^2} A(t) \sin (\omega_L t + \phi_\text{PM}) \text{,} \\
\label{eq:amplitude-quadrature}
\langle I_{I}(t) \rangle & = B \Delta_\text{pc} A(t) \sin (\omega_L t +\phi_\text{AM}) \text{,}
\end{align}
\end{subequations}
with the relative phase between these responses $\tan (\phi_\text{PM} - \phi_\text{AM}) = \omega_L / \kappa$, and the common amplitude scale
\begin{equation}
\label{eq:quadrature-scale}
B = \sqrt{ \frac{2 \epsilon P_\text{LO} \hbar \omega_p \bar{n} \kappa g_c ^2 }{(\kappa^2 + \Delta_\text{pc}^2 + \omega_L^2)^2 - 4 \Delta_\text{pc}^2 \omega_L^2} }\text{.}
\end{equation}

This result shows that, for a detuned cavity probe, the oscillator's dynamics are encoded in both the amplitude and phase quadrature of the heterodyne signal, but with different phase.  This phase difference prevents extraction of all signal information by a simple linear combination of the quadratures, so we instead perform a simultaneous fit to both.

\begin{figure}[!tb]
	\centering
	\includegraphics{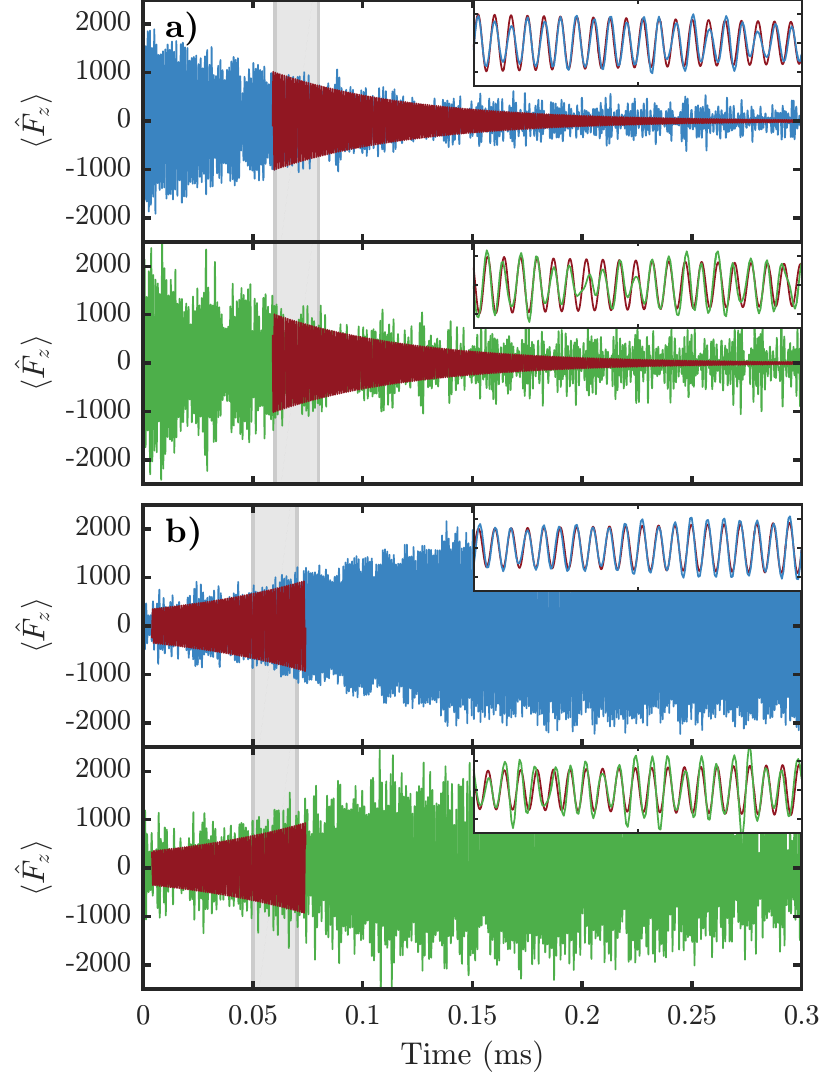}
	\caption{
		\label{fig:timedomain-quadratures} Estimates of the spin dynamics for a \textbf{(a)} damped and \textbf{(b)} amplified spin oscillator, obtained from both the amplitude (green, lower) and phase (blue, upper) quadratures of the demodulated heterodyne signal.  This shows the entire signal and simultaneous fit for the results shown in Fig. 2 of the main text.
	}
\end{figure}

The fits in the paper are performed by assuming $A(t)=A_0 e^{-i \Gamma t/2}$, and simultaneously fitting Equations.\ \ref{eq:phase-quadrature} and\ \ref{eq:amplitude-quadrature} directly to both demodulated quadratures, allowing the initial amplitude $A_0$, exponential damping rate $\Gamma$, shifted response frequency $\omega_L^\prime$ and a common arbitrary phase $\phi$ to vary.  Since the relative signal to noise of each quadrature varies with detuning, this naturally provides the appropriate weighting of each quadrature in the fit.  Because the oscillator response and cavity coupling become inherently non-linear for large amplitude oscillations, these solutions and the choice of an exponential model function are only valid for small excitations.  To consistently extract the damping (amplification) rates for small amplitude oscillations across various conditions, we iteratively scan the start (stop) time of the fit window, choosing the longest time range within which the amplitude of the fit result remains below a fixed threshold, $F_z(t) < 1000$.

\section*{S.3 Cavity cooling and sideband asymmetry thermometry}	

The equilibrium spin temperature for a damped spin oscillator is solely determined by the quantum fluctuations of the cavity photon number at the Larmor frequency.  For a spin precessing around a large magnetic field along the $\mathbf{x}$ axis, the eigenstates have well-defined angular momentum $m$ along this axis.  If the spin experiences a weak coupling to the cavity field of the form $\mathcal{H} = \hbar g_c \hat{n} \hat{F}_z = \hbar g_c \hat{n} (\hat{F}_{+} + \hat{F}_{-})/2$, the photon fluctuations mediate transitions adding or removing units of angular momentum along $\mathbf{x}$.  The transition rate for a collective spin $F$ from state $|m\rangle$ to $|m+1\rangle$ is given by
\begin{equation}
\Gamma_{\uparrow} = P_m \frac{g_c^2}{4} |\langle m+1 | \hat{F}_{+} | m \rangle|^2 S_{nn}(-\omega_L)\text{,}
\end{equation}
where $P_m$ is the population in state $m$, and $S_{nn}(\omega)$ is the power spectral density of photon number fluctuations inside the cavity.  The rate of the reverse process is
\begin{equation}
\Gamma_{\downarrow} = P_{m+1} \frac{g_c^2}{4} |\langle m | \hat{F}_{-} | m+1 \rangle|^2 S_{nn}(\omega_L)\text{.}
\end{equation}
Since this is the only process mediating energy exchange with the spin ensemble, in equilibrium, detailed balance requires these rates to be equal, and $P_{m+1}=P_{m} \exp (-\hbar \omega_L / k_B T)$, which implies
\begin{equation}
\label{eq:spin-equilibrium}
\frac{ S_{nn} (\omega_L) } { S_{nn}(-\omega_L) } = \exp \frac{ \hbar \omega_L  }{ k_B T }\text{.}
\end{equation}
This result is also a statement of the effective temperature of the photon shot-noise in the cavity mode at the Larmor frequency, showing that the spin equilibrates to the same temperature.

For damping by a single coherent tone, one finds that the equilibrium temperature is independent of the intra-cavity intensity.  However, the photon shot-noise from the second, on-resonance tone causes additional heating, with the time-averaged photon shot-noise spectrum given by
\begin{equation}
S_{nn}(\omega) = \frac{2 \kappa \bar{n}_\text{damp}}{(\omega + \Delta_\text{pc})^2 + \kappa^2} + \frac{2 \kappa \bar{n}_\text{probe}}{\omega^2 + \kappa^2}\text{,}
\end{equation}
where $\Delta_\text{pc}$ is the detuning of the damping tone which has mean intra-cavity photon number $\bar{n}_\text{damp}$, and $\bar{n}_\text{probe}$ is the corresponding intensity of the probe.  Due to the additional heating of this second tone, the resulting equilibrium depends on the balance between the intensity of the two tones.  We choose this balance to achieve a fixed measurement cooperativity, defined by 
\begin{equation}
C_s= \frac{ 4 g_s^2 \bar{n}_\text{probe} } {\kappa \Gamma_\text{opt}}\text{,}
\end{equation}
where $\Gamma_{opt}=g_s^2 \bigl[S_{nn}(\omega_L) - S_{nn}(-\omega_L) \bigr]$ is the optical damping due to the off-resonance tone.  The detuning that minimizes the resulting temperature is given by $\Delta_\text{pc} = \beta \sqrt{\omega_L^2 + \kappa^2}$, which yields a minimum equilibrium temperature of
\begin{equation}
\exp \frac{ \hbar \omega_L  }{ k_B T } = 1 + 2 \Biggl[ 
\sqrt{1+\frac{\kappa^2}{\omega_L^2}} + \frac{C_s \kappa^2 } {\kappa^2 + \omega_L^2} - 1
 \Biggr]^{-1}\text{.}
\end{equation}
The average measurement cooperativity for the data shown in Fig. 4c of the main text was $C_s = 1.6$.

\end{document}